\documentclass[12pt]{article}
\input{epsf.sty}
\def\mydate{4 October 2002}
\def\ignore#1{{}}

\let\oldtheequation=\theequation
\def\doteqs#1{\setcounter{equation}{0}
            \def\theequation{{#1}.\oldtheequation}}
\newcounter{sxn}
\def\sx#1{\addtocounter{sxn}{1} \vskip 1.cm  \goodbreak
\noindent{\large\bf\leftline{\thesxn.~~#1}} \nobreak \vskip -.5cm}
\def\sxn#1{\sx{#1} \doteqs{\thesxn}}

\newcounter{axn}

\date{}

\newdimen\mybaselineskip
\renewcommand{\baselinestretch}{1.25}

\newcommand{\beeq}{\begin{equation}}
\newcommand{\eneq}{\end{equation}}
\newcommand{\beqn}{\begin{eqnarray}}
\newcommand{\eeqn}{\end{eqnarray}}

\def\dd{\partial}
\def\la{\raise.16ex\hbox{$\langle$}\lower.16ex\hbox{}  }
\def\ra{\, \raise.16ex\hbox{$\rangle$}\lower.16ex\hbox{} }
\def\go{\rightarrow}

\def\onehalf{ \hbox{${1\over 2}$} }

\def\ep{\epsilon}

\begin{document}
\thispagestyle{empty}

\baselineskip=12pt

{\small \noindent \mydate  \hfill OU-HET 419/2002}

 {\small \noindent (Corrected. 5 December 2002) \hfill }

\baselineskip=40pt plus 1pt minus 1pt

\vskip 4cm

\begin{center}

{\LARGE \bf Weyl Invariant Spacetime}\\

\vspace{3.0cm}
\baselineskip=20pt plus 1pt minus 1pt

{\bf  Yutaka Hosotani}\\
\vspace{.1cm}
{\it Department of Physics, Osaka University,
Toyonaka, Osaka 560-0043, Japan}\\
\end{center}

\vskip 2.cm
\baselineskip=20pt plus 1pt minus 1pt

\begin{abstract}
A new class of  Weyl invariant backgrounds are presented in terms of 
the metric $G_{\mu\nu}$ and the anti-symmetric Kalb-Ramond fields
$B_{\mu\nu}$. The ten-dimensional spacetime is a product of four-dimensional
flat spacetime and curved six-dimensional spacetime having
nonvanishing Ricci tensors.  The non-vanishing Kalb-Ramond  field strengths
cannot be written globally as $H=dB$,   being of the
monopole type.  Nevertheless they define homogeneous spacetime with no
singularity.
\end{abstract}


\newpage


\newpage

\sxn{Introduction }

In the string theory the background spacetime cannot be arbitrarily 
assigned.   For the consistency of the theory the Weyl invariance 
has to be maintained.  At  low energies the spacetime is described
in terms of the metric $G_{\mu\nu}$, the anti-symmetric  Kalb-Ramond fields
$B_{\mu\nu}$, and the dilaton field $\Phi$.  In the background
the trace of world-sheet energy momentum tensors is cast
in the form
\beeq
{T^a}_a = 
- {1\over 2 \alpha'} \beta^G_{\mu\nu} g^{ab} \dd_a X^\mu \dd_b X^\nu
- {i\over 2 \alpha'} \beta^B_{\mu\nu} \ep^{ab} \dd_a X^\mu \dd_b X^\nu
- {1\over 2} \beta^\Phi R  ~~,
\label{Weyl1}
\eneq
which must vanish to maintain the Weyl invariance.  Here $1/(2\pi \alpha')$ and
$g_{ab}$ are the  string tension and the world-sheet metric. $\ep^{ab}$ is the 
two-dimensional antisymmetric tensor.
$R$ is the scalar curvature of the spacetime. 
Callan et al.\ \cite{Callan1} have shown that 
the condition for having  vanishing beta functions in the critical dimension
$D=D_c$ leads to 
\beqn
&&\hskip -1cm
\beta^G_{\mu\nu} = \alpha' \bigg\{ 
(Ricci)_{\mu\nu} + 2 \nabla_\mu \nabla_\nu \Phi
- {1\over 4} H_{\mu\rho\sigma} {H_\nu}^{\rho\sigma} \bigg\} 
+ O (\alpha'^2) = 0~, \cr
\noalign{\kern 10pt}
&&\hskip -1cm
\beta^B_{\mu\nu} = \alpha' \bigg\{ 
- {1\over 2} \nabla^\rho H_{\rho\mu\nu} 
+ \nabla^\rho \Phi \cdot  H_{\rho\mu\nu} \bigg\}
+ O (\alpha'^2) = 0 ~, \cr
\noalign{\kern 10pt}
&&\hskip -1cm
\beta^\Phi  = \alpha' \bigg\{
- {1\over 2} \nabla^2 \Phi + \nabla_\mu \Phi \nabla^\mu \Phi  
- {1\over 24} \, H_{\mu\nu\lambda}  H^{\mu\nu\lambda} \bigg\} 
+ O (\alpha'^2) =0 ~.
\label{Weyl2}
\eeqn
This set of the equations can be derived  from the 
action
\beqn
&&\hskip -1cm
I = {1\over 2 \kappa^2}\int d^D x \, | \det G |^{1/2} ~ {\cal L} \cr
\noalign{\kern 10pt}
&&\hskip -1cm
{\cal L} = e^{-2\Phi} \Bigg\{
R - {1\over 12} \, H_{\mu\nu\lambda}  H^{\mu\nu\lambda}
+ 4 \nabla_\mu \Phi \nabla^\mu \Phi + O(\alpha')  \Bigg\} 
\label{Weyl3}
\eeqn
by taking variation over $G_{\mu\nu}$, $B_{\mu\nu}$ and $\Phi$.
In other words Weyl invariant backgrounds are classical solutions 
to the equations of motion in the field theory defined by (\ref{Weyl3}).

Most of the backgrounds considered in the literature are Ricci flat,
$(Ricci)_{\mu\nu} = 0$,  and have vanishing $H_{\mu\nu\lambda}$'s.
It is of great interest to know if there are other kinds of backgrounds
with nonvanishing $(Ricci)_{\mu\nu}$ and $H_{\mu\nu\lambda}$.\cite{Braaten}   Here
we loosen the conditions for the consistency.  In the string
theory it is assumed that the extra six dimensions are all spacelike, i.e.\ 
the signature of the metric is all positive.
We drop this restriction.   We ask if there are new type of solutions to
(\ref{Weyl2}).  Spacetime satisfying (\ref{Weyl2}) 
is said, in the present paper,  to be  Weyl invariant.

At the bottom of this investigation is the observation that the Kalb-Ramond
fields $B_{\mu\nu}$ are conformally invariant at the classical level and 
 have natural configurations in six dimensions.\cite{Kalb}
It is a remarkable fact that the string theory produces the Kalb-Ramond 
fields  which may account for why we can see only four dimensions
at very low energies.  We shall see below that in one of the examples 
extra dimensions are compact and there appears high symmetry
between the spacelike and timelike dimensions.

\sxn{Consistency conditions}

We shall impose another constraint to set up the analysis.  We restrict
ourselves to configurations with a constant dilaton field
$\Phi =$constant.  It seems very difficult to have configurations
with spatially varying $\Phi$ when extra-dimensional spacetime is compact.

With this restriction imposed the consistency conditions (\ref{Weyl2}) reduce,
to $O(\alpha')$,  to
\beqn
&&\hskip -1cm
(Ricci)_{\mu\nu} =
 {1\over 4} H_{\mu\rho\sigma} {H_\nu}^{\rho\sigma}  ~, \cr
\noalign{\kern 10pt}
&&\hskip -1cm
 \nabla_\rho H^{\rho\mu\nu} = 
{1\over \sqrt{|G|}} ~ \dd_\rho \bigg\{ \sqrt{|G|} ~ H^{\rho\mu\nu} \bigg\}
= 0 ~,  \cr
\noalign{\kern 10pt}
&&\hskip -1cm
R  = 0 ~~,~~
 H_{\mu\nu\lambda}  H^{\mu\nu\lambda}  =0 ~.
\label{Weyl4}
\eeqn
The last equality implies that, if the extra dimensions are all spacelike,
i.e.\ the signature of the metric is all positive, 
all components of $H_{\mu\nu\lambda}$'s vanish so that the resultant
space is Ricci flat.  We shall consider more general spacetime.

The metric of six dimensional spacetime is written as 
\beeq
ds^2 =  \eta_{ab} ~ e^a \otimes e^b
\label{metric1}
\eneq
where $e^a$'s ($a=1 \sim 6$) are tetrad 1-forms. $\eta_{ab}$ 
is a diagonal matrix, whose elements are either $+1$ or $-1$.
In all of the Weyl invariant spacetime discussed bellow
$\det \, (\eta_{ab}) $ is $-1$.   The spin connection 
1-forms are defined by $d e^a + {\omega^a}_b \wedge e^b = 0$.
The curvature 2-forms are then given by 
${\cal R}_{ab} = d\omega_{ab} + {\omega_a}^c \wedge \omega_{cb}$,
whose components are Riemann tensors; 
${\cal R}_{ab} = \onehalf R_{abcd} ~ e^c \wedge e^d$. 
Ricci tensors are $(Ricci)_{ab} = {R_{acb}}^c$.

The Kalb-Ramond  fields define  2-form fields; 
$B = \onehalf B_{\mu\nu} ~ dx^\mu \wedge dx^\nu 
=  \onehalf B_{ab} ~e^a \wedge e^b$.  Their field strengths define
3-form fields $H = {1\over 6} H_{abc} ~ e^a \wedge e^b \wedge e^c = d B$. 
The equations in (\ref{Weyl4}) are rewritten as
\beqn
&&\hskip -1cm
(Ricci)_{ab} = {1\over 4} \, H_{acd} {H_b}^{cd} \equiv S_{ab} ~~,  
\label{cond1} \\
\noalign{\kern 10pt}
&&\hskip -1cm
R = 0 ~~,  
\label{cond2} \\
\noalign{\kern 10pt}
&&\hskip -1cm
dH = 0 ~~,~~ {}^* d ( {}^* H ) = 0 ~~.
\label{cond3}
\eeqn
${}^*$ denotes Hodge dual.  We note that the $B$-fields are gauge fields
for closed strings.  The theory is invariant under a gauge transformation
$B \go B + d \Lambda$ where $\Lambda$ is a 1-form gauge function.
Just like the monopole fields in the electrodynamics in three spatial
dimensions, the potential $B$ need not be defined globally.  The field
strengths $H$  must satisfy (\ref{cond3}).

\sxn{Example I. $SU(2) \otimes \overline{SU(2)}$}

A nontrivial Weyl-invariant background is obtained by considering
a product of two three-spheres ($S^3$).  Let us first recall some properties of 
a manifold $S^3$.   $S^3$ is isomorphic to a group manifold $SU(2)$.
The relationship is most easily established by the correspondence 
\beeq
S^3 : ~ y_1^2 + y_2^2 + y_3^2 + y_4^2 = 1 
~\Leftrightarrow~
 \Omega = y_4 + i \vec y \cdot \vec \tau 
\in SU(2) 
\label{correspondence}
\eneq
where $\tau^j$'s ($j=1,2,3$) are Pauli matrices.  Maurer-Cartan 
1-forms of $SU(2)$, 
$\sigma^j = -i {\rm Tr}\,  \tau^j \Omega^{-1} d \Omega$ satisfy
$d\sigma^j = \onehalf \ep^{jkl} \sigma^k \wedge \sigma^l$.  
In terms of $\sigma^j$'s the metric of $S^3$ or $SU(2)$ is given by
$ds^2 = \sum_{j=1}^3 \sigma^j \otimes \sigma^j$.

Now we prepare two sets of $SU(2)$;
\beeq
d\sigma^j = \onehalf \ep^{jkl} \sigma^k \wedge \sigma^l
~~,~~
d\rho^j = \onehalf \ep^{jkl} \rho^k \wedge \rho^l ~~.
\label{structure1}
\eneq
The metric of the six-dimensional manifold,
$SU(2) \otimes \overline{SU(2)}$,  is written as
\beeq
ds^2 = \sum_{j=1}^3 (\sigma^j \otimes \sigma^j 
                     - \rho^j \otimes \rho^j ) ~~, 
\label{metric2}
\eneq
or equivalently,
\beqn
&&\hskip -1cm
e^1 = \sigma^1 ~,~ e^2 = \sigma^2 ~,~ e^3 = \sigma^3 ~,~
e^4 = \rho^1 ~,~ e^5 = \rho^2 ~,~ e^6 = \rho^3 ~,~ \cr
\noalign{\kern 10pt}
&&\hskip -1cm
\eta_{ab} = diag (1,1,1,-1,-1,-1)   ~~.
\label{metric3}
\eeqn
Curvature 2-forms are given by
\beqn
&&\hskip -1cm
{\cal R}_{j,k} = {1\over 4} \, e_j \wedge e_k \cr
\noalign{\kern 5pt}
&&\hskip -1cm
{\cal R}_{j+3,k+3} = - {1\over 4} \, e_{j+3} \wedge e_{k+3} \cr
\noalign{\kern 10pt}
&&\hskip -1cm
{\cal R}_{j,k+3} = 0~~,~~ (j,k = 1,2,3) ~.
\label{curvature1}
\eeqn
Ricci tensors are found to be
\beeq
(Ricci)_{ab} = {1\over 2} ~ \delta_{ab} 
= {1\over 2} \,  diag (1,1,1,1,1,1)  ~~.
\label{Ricci1}
\eneq
The scalar curvature $R = {(Ricci)_a}^a$ vanishes.
The length scale, $L$, in the metric has been set to be unity.
Every quantity scales with $L$ with an appropriate power.

We need to find 3-form field strengths $H$ satisfying (\ref{cond1})
and (\ref{cond3}).  In this connection  it is very fruitful to recall how special
configurations in non-Abelian gauge theory in four dimensions arise.
They  might have  been realized in the early universe.  If the universe is described
by the Robertson-Walker metric with $k=+1$, namely with a spatial
section $S^3$,  then there appears a natural mapping in
the configurations of $SU(2)$ gauge fields.  The 1-form gauge potential
takes the form $A = f(t) ~ \vec \sigma \cdot \vec \tau$
where $\sigma^k$'s are the Maurer-Cartan forms.  This ansatz
simultaneously solves the Einstein equations and Yang-Mills
equations.\cite{Hosotani1,Hosoya}  The resultant space is homogeneous and
isotropic.   Its generalization solves the equations  in the
Einstein-Weinberg-Salam theory as well. The homogeneity of the space is maintained,
but the space necessarily becomes anisotropic due to the $U(1)$ gauge
interaction.\cite{Hosotani2}  A crucial point is that 1-form gauge potentials
have natural configurations on $SU(2) \sim S^3$.

In the problem under discussions we have the Abelian antisymmetric tensor
fields $B$, namely 2-form fields.   To our surprise there are natural
configurations on the manifold (\ref{metric2}).  First notice that
the 3-form field strength $H=d(\sigma^j \wedge \rho^k)$ is self-dual,
${}^* H = + H$, as straightforward calculations show.  
It therefore automatically solves $d({}^* H) =dH= 0$. The symmetrized 
version 
\beeq
B_0 = \sum_{j=1}^3 \sum_{k=1}^3 \sigma^j \wedge \rho^k ~~,~~
H_0 = dB_0
\label{conf1}
\eneq
yields
\beqn
&&\hskip -1cm
dH_0 = 0 ~~,~~ {}^* d ( {}^* H_0 ) = 0 ~~, \cr
\noalign{\kern 10pt}
&&\hskip -1cm
(S_0)_{ab} = 
{1\over 4} \, (H_0)_{acd} {(H_0)_b}^{cd} = -{3\over 2} \delta_{ab} ~~.
\label{conf2}
\eeqn
The configuration $H_0$ solves almost all of the consistency equations except  it
gives  a wrong sign for $S_{ab}$. 

We next consider the following self-dual or anti-self-dual configuration;
\beqn
&&\hskip -1cm
H_\pm = \sigma^1 \wedge \sigma^2 \wedge \sigma^3
  \mp \rho^1 \wedge \rho^2 \wedge \rho^3  \cr 
\noalign{\kern 10pt}
&&\hskip -.3cm
= e^1 \wedge e^2 \wedge e^3  \mp e^4 \wedge e^5 \wedge e^6  \cr 
\noalign{\kern 10pt}
&&\hskip -1.2cm
{}^* H_\pm = \pm H_\pm ~~.
\label{conf3}
\eeqn
It follows from (\ref{structure1}) that $dH_\pm = 0$.  We have
\beqn
&&\hskip -1cm
dH_\pm = 0 ~~,~~ {}^* d ( {}^* H_\pm ) = 0 ~~, \cr
\noalign{\kern 10pt}
&&\hskip -1cm
(S_\pm)_{ab} = 
{1\over 4} \, (H_\pm)_{acd} {(H_\pm)_b}^{cd} 
= {1\over 2} \delta_{ab} ~~.
\label{conf4}
\eeqn
$H_\pm$ itself solves (\ref{cond1}) and (\ref{cond3}).  The configuration
$H_\pm$ cannot be written  globally as $H_\pm = d B_\pm$.  Nevertheless
it is a legitimate physical configuration as it solves the equations
of motion (\ref{cond3}).  Its coupling to matter, say, strings, is well
defined thanks to the gauge invariance associated with $B$.   It is like a
monopole configuration in  the four-dimensional
electromagnetism.\cite{Orland}  The vector potential for a magnetic monopole
cannot be written globally without introducing a Dirac string.\cite{Dirac} 
We remark that the field strength
$H_\pm$ is regular everywhere.  It gives homogeneous spacetime. 

General solutions to (\ref{cond1}) and (\ref{cond3}) in the metric 
(\ref{metric2}) are
\beqn
(i)&& H = \pm H_+ \cr
\noalign{\kern 10pt}
(ii)&& H = \alpha H_- + \beta H_0 ~~,~~ \alpha^2 - 3 \beta^2 =1 ~~.
\label{general1}
\eeqn
As $H_0$ is self-dual, the interference term occurring in $S_{ab}$
for the case $(ii)$ vanishes.  (\ref{metric2}) and 
(\ref{general1}) define Weyl invariant spacetime.

The (anti-)self-duality relation for $H$ has played a vital role.
In six dimensions self-dual or anti-self-dual configurations are
possible  only when there are an odd number of timelike dimensions.

\sxn{Example II. $SL(2,R) \otimes \overline{SL(2,R)}$}

The second example of Weyl invariant spacetime is obtained by 
replacing $SU(2)$ by $SL(2,R)$ in the previous example.  Generators 
of $SL(2,R)$ satisfy
\beeq
[T_1, T_2] = T_3 ~~,~~ [T_2, T_3] = T_1 ~~,~~ [T_3, T_1] = - T_2 ~~.
\label{structure2}
\eneq
The corresponding Maurer-Cartan 1-forms satisfy
\beeq
d\sigma^1 = \sigma^2 \wedge \sigma^3 ~~,~~ 
d\sigma^2 = - \sigma^3 \wedge \sigma^1 ~~,~~ 
d\sigma^3 = \sigma^1 \wedge \sigma^2 ~~.
\label{structure3}
\eneq 
The metric of the manifold $SL(2,R)$ is given by 
$ds^2 = \sigma^1 \otimes \sigma^1 - \sigma^2 \otimes \sigma^2 
  + \sigma^3 \otimes \sigma^3 $.

We prepare two sets of $SL(2,R)$ whose Maurer-Cartan 1-forms are denoted by 
$\{ \sigma^j \}$ and $\{ \rho^j \}$.
The metric $ds^2 = \eta_{ab} e^a \otimes e^b$ of the six-dimensional manifold,
$SL(2,R) \otimes \overline{SL(2,R)}$,  is given by
$e^j = \sigma^j ~,~ e^{j+3} = \rho^j$ ($j=1,2,3$) and 
\beeq
\eta_{ab} = diag \, (1,-1,1,-1,1,-1)  ~~.
\label{metric6}
\eneq
Curvature 2-forms are given by
\beqn
&&\hskip -1cm
{\cal R}_{j,k} = -{1\over 4} \, e_j \wedge e_k \cr
\noalign{\kern 5pt}
&&\hskip -1cm
{\cal R}_{j+3,k+3} = + {1\over 4} \, e_{j+3} \wedge e_{k+3} \cr
\noalign{\kern 10pt}
&&\hskip -1cm
{\cal R}_{j,k+3} = 0~~,~~ (j,k = 1,2,3) ~.
\label{curvature2}
\eeqn
Compared with (\ref{curvature1}) in the previous example, the sign 
is opposite.  Ricci tensors become
\beeq
(Ricci)_{ab} 
= - {1\over 2} \, diag \, (1,-1,1,1,-1,1) ~~.
\label{Ricci2}
\eneq
The scalar curvature $R$ vanishes.

The configurations solving (\ref{cond1}) and (\ref{cond3}) are
\beqn
&&\hskip -.9cm
H ~  =  \pm \,  H_\pm ~~, \cr
\noalign{\kern 10pt}
&&\hskip -1.cm
H_\pm  =  \sigma^1 \wedge \sigma^2 \wedge \sigma^3
  \pm \rho^1 \wedge \rho^2 \wedge \rho^3   \cr 
\noalign{\kern 10pt}
&&\hskip -.3cm
=  e^1 \wedge e^2 \wedge e^3  \pm e^4 \wedge e^5 \wedge e^6   \cr 
\noalign{\kern 10pt}
&&\hskip -1.2cm
{}^* H_\pm = \pm H_\pm ~~.
\label{conf5}
\eeqn
Again these $H$'s cannot be written globally as $H = dB$.  $S_{ab}$ 
for these configurations is exactly given by the right hand side of
(\ref{Ricci2}), solving (\ref{cond1}).

The gauge potential $B^{(jk)} = \sigma^j \wedge \rho^k$ yields
anti-self-dual field strength; ${}^* (d B^{(jk)}) = - d B^{(jk)}$.
However, a configuration of the type $B_0 = \sum_{j=1}^3 \sum_{k=1}^3
 B^{(jk)}$ does not solve (\ref{cond1}).  On the manifold 
$SL(2,R) \otimes \overline{SL(2,R)}$ the allowed  $H$ configurations
are (\ref{conf5}) only.  We remark that the manifold $SO(2,2)$ is 
isomorphic to $SL(2,R) \otimes \overline{SL(2,R)}$.

\sxn{Example III. $SU(2) \otimes SL(2,R)$ and 
$\overline{SU(2)} \otimes \overline{SL(2,R)}$}

The third example is given by a product of $SU(2)$ and 
$SL(2,R)$.   Let $\{ \sigma^j \}$ and $\{ \rho^j \}$ be
Maurer-Cartan forms of $SU(2)$ and  $SL(2,R)$, respectively. 
The metric is given by 
\beeq
ds^2 = \sigma^1 \otimes \sigma^1 +  \sigma^2 \otimes \sigma^2
 + \sigma^3 \otimes \sigma^3 + \rho^1 \otimes \rho^1 
  - \rho^2 \otimes \rho^2 + \rho^3 \otimes \rho^3 ~, 
\label{metric7}
\eneq
or equivalently  $e^j = \sigma^j ~,~ e^{j+3} = \rho^j$ ($j=1,2,3$) and
$\eta_{ab} = diag (1,1,1,1,-1, 1)$.  The curvature 
2-forms are given by (\ref{curvature1}) so that
\beeq
(Ricci)_{ab} = \onehalf ~ diag(1,1,1,-1,1, -1) ~,
\label{curvature3}
\eneq
yielding a vanishing scalar curvature.   The Kalb-Ramond field
strengths are 
\beeq
H = H_\pm = 
 e^1 \wedge e^2 \wedge e^3  \mp e^4 \wedge e^5 \wedge e^6 
= \pm {}^* H_\pm ~.
\label{conf6}
\eneq
Eq.\ (\ref{cond1}) is satisfied as
\beeq
S_{ab} =  \onehalf ~ diag(1,1,1,-1,1, -1) ~.
\label{conf7}
\eneq
This time again  $B^{(jk)} = \sigma^j \wedge \rho^k$ yields
anti-self-dual field strengths; ${}^* (d B^{(jk)}) = - d B^{(jk)}$.

The fourth example is obtained by flipping the signature of the third
example, namely by $\overline{SU(2)} \otimes \overline{SL(2,R)}$.
The curvature 2-forms are given by (\ref{curvature2}), whereas the Ricci
tensors turn out the same as (\ref{curvature3}).  
The Kalb-Ramond field strengths must be
$H = H_\pm = 
 e^1 \wedge e^2 \wedge e^3  \pm e^4 \wedge e^5 \wedge e^6 $.
$B^{(jk)} = \sigma^j \wedge \rho^k$ yields self-dual field 
strengths on the manifold; ${}^* (d B^{(jk)}) = + d B^{(jk)}$.

\sxn{Other group manifolds:  $SO(4)$ and $SO(3,1)$}

It may be worth clarifying why the manifolds $SO(4)$ and $SO(3,1)$
are not Weyl invariant.  The manifold $SO(4)$ is locally isomorphic to 
$SU(2) \otimes SU(2)$.  The metric is given, in terms of the $\sigma^j$'s
and  $\rho^j$'s in Section 3,  by $ds^2 = \sum_{j=1}^3 (\sigma^j \otimes
\sigma^j   +\rho^j \otimes \rho^j )$  instead of (\ref{metric2}), which
leads to (\ref{Ricci1}) and  nonvanishing scalar curvature $R=3$.  
Hence the condition (\ref{cond2}) cannot be satisfied.  A varying 
dilaton field $\Phi(x)$ may give a nontrivial solution to (\ref{Weyl2}).
It, however, is  difficult to have such $\Phi$ on a compact manifold.

The situation is similar for the manifold $SO(3,1)$.  This manifold 
is of special interest.  If the ten-dimensional spacetime is a product
of four-dimensional Minkowski spacetime and six-dimensional $SO(3,1)$,
then Lorentz symmetry transformations of the four-dimensional Minkowski spacetime
have one-to-one correspondence to points or translations in the extra
six-dimensional spacetime.  It could be the origin of the local
Lorentz invariance of the four-dimensional spacetime, when the
four-dimensional spacetime is promoted to curved spacetime.

The manifold $SO(3,1)$ is  isomorphic to $SL(2,C)$ whose complex
Maurer-Cartan 1-forms are denoted by $\sigma^k + i \rho^k$ ($k=1,2,3$).
They satisfy
\beeq
d(\sigma^j + i \rho^j)
= \onehalf \ep^{jkl} (\sigma^k + i \rho^k) \wedge 
(\sigma^l + i \rho^l) ~~.
\label{structure3}
\eneq
The metric is 
\beeq
ds^2 = \sum_{j=1}^3 (\sigma^j \otimes \sigma^j 
                     - \rho^j \otimes \rho^j ) ~~. 
\label{metric8}
\eneq
Although it has the same form as (\ref{metric2}), its content is quite
different.  With tetrads $e^j = \sigma^j, e^{j+3} = \rho^j$ ($j=1,2,3$)
curvature 2-forms are given by
\beqn
&&\hskip -1cm
{\cal R}_{j,k} = -{\cal R}_{j+3,k+3} 
= {1\over 4} \, ( e^j \wedge e^k - e^{j+3} \wedge e^{k+3} ) \cr
\noalign{\kern 5pt}
&&\hskip -1cm
{\cal R}_{j,k+3} 
= - {1\over 4} \, ( e^j \wedge e^{k+3} - e^{k} \wedge e^{j+3} ) 
~~,~~ (j,k  = 1,2,3) ~.
\label{curvature4}
\eeqn
Ricci tensors are
\beeq
(Ricci)_{ab} = \eta_{ab} 
= diag \, (1,1,1,-1,-1, -1) ~.
\label{Ricci3}
\eneq
The manifold $SO(3,1)$ is maximally symmetric.  
Its scalar curvature is non-vanishing. 

The manifold is not compact.  There may be configurations with
$x$-dependent $\Phi$ which solve the equations in (\ref{Weyl2}).
The unboundedness of the manifold make it possible to have such
configurations, though we have not so far found one.  In Weyl invariant
spacetime the Kalb-Ramond field strengths $H \sim H_\alpha$ or $H_\beta$
where
\beqn
&&\hskip -1cm
H_\alpha = + e^1 \wedge e^2 \wedge e^3 - e^1 \wedge e^5 \wedge e^6
 + e^2 \wedge e^4 \wedge e^6  - e^3 \wedge e^4 \wedge e^5 ~, \cr
\noalign{\kern 5pt}
&&\hskip -1cm
H_\beta = - e^4 \wedge e^5 \wedge e^6   + e^2 \wedge e^3 \wedge e^4
 - e^1 \wedge e^3 \wedge e^5 + e^1 \wedge e^2 \wedge e^6
\label{conf8}
\eeqn
would play an important role.  $H_\alpha$ and $H_\beta$ satisfy
\beqn
&&\hskip -1cm
d H_\alpha = d H_\beta = 0 ~~, \cr
\noalign{\kern 5pt}
&&\hskip -1cm
{}^* H_\alpha = H_\beta ~~,~~ {}^* H_\beta = H_\alpha ~~,  \cr
\noalign{\kern 5pt} 
&&\hskip -1cm
(S_\alpha)_{ab} = + \eta_{ab} ~~,~~ (S_\beta)_{ab} = -  \eta_{ab} ~~.
\label{conf9}
\eeqn

\sxn{Discussions}

In this paper we have investigated  Weyl invariant
spacetime  defined as  a background solving the set of the equations
in (\ref{Weyl2}).  Beside the well-known Ricci flat spacetime
we have found a new class of backgrounds with nonvanishing 
Ricci tensors.  The Kalb-Ramond  field strengths assume 
nontrivial configurations, balancing the two sides of the Einstein
equations.  The scalar curvature vanishes as the dilaton field is
set to be constant.

The construction of Weyl invariant spacetime is based on the fact
that the Kalb-Ramond field strengths $H$ are 3-form fields satisfying
$dH = {}^* d ({}^* H) = 0$.  We prepare a six-dimensional manifold
in the form of a product of two three-dimensional group manifolds.
On each three-dimensional group manifold,  $H=e^1 \wedge e^2 \wedge e^3$ 
where $e^j$'s are Maurer-Cartan forms solves the equations for $H$.
We need two group manifolds so as to have  vanishing scalar
curvature in  six dimensions.  One of the two group 
manifolds must have negative scalar curvature, which is achieved either by
preparing a group manifold with  intrinsically negative 
curvature or by flipping the signature in the metric.

In either case there have appeared timelike dimensions which makes it
difficult to apply our results to, say, the string theory.  Excitations
in timelike dimensions would cause instability.  It has to be explored
if there is a way to control excitations in timelike dimensions.
Many arguments have been put forward in the literature that extra timelike
dimensions  can physically make sense when certain conditions are 
met.\cite{Sakharov}-\cite{Matsuda}  Another related point to be clarified is about
the existence of  supersymmetry on the Weyl invariant manifolds.  

In the examples discussed above the Kalb-Ramond field strengths have
constant magnitude.  They are of the monopole type in the 
sense that they cannot be written globally as $H = dB$.  
What would be dynamics of strings on such backgrounds?  It is about 
 strings on group manifolds.\cite{Witten}
Motion of particles in a constant magnetic field $F=dA$ or of open strings 
in a constant $B$ effectively produces non-commutative geometry.
What would be dynamics of closed strings in a constant $H$ on a group
manifold?  

The first example of the Weyl invariant spacetime, 
$SU(2) \otimes \overline{SU(2)}$, has many interesting
properties.   First of all the manifold is compact.  Secondly
there is one parameter family of Weyl invariant spacetime
on this manifold.  $H$ is given by (\ref{general1}).   It is interesting
to know the relevance of  the  parameter $\beta$ in (\ref{general1}).   The
configuration
$H_0$ in (\ref{conf1}) bridges the two $SU(2)$ manifolds.  Thirdly
there is accidental symmetry of the manifold, namely symmetry under
the interchange of the two $SU(2)$'s.  Fourthly a six-dimensional flat spacetime
with three timelike dimensions has Majorana-Weyl spinors.  

Motivated by the string theory, we have investigated Weyl invariant
spacetime in general framework. 
It would be of great interest  to know if the Weyl invariant spacetime
discussed in the present paper has applications in physics.   We hope
to come back to this point in near future.

\vskip 1.cm

\leftline{\bf Acknowledgments}

I benefited from  enlightening  discussions with  many of my colleagues
including  K.\ Higashijima, T.\ Nakatsu, T.\ Ohnishi, N.\ Ohta, M.\ Sato,  T.\
Tanaka, and K.\ Tsuda.  I would like to duly thank them.  This work was supported in
part by  Scientific Grants  from the Ministry of Education and Science,
Grant No.\ 13135215 
 and Grant No.\ 13640284.

\vskip .5cm

\def\jnl#1#2#3#4{{#1}{\bf #2} (#4) #3}

\def\Zphys{{\em Z.\ Phys.} }
\def\jssc{{\em J.\ Solid State Chem.\ }}
\def\jpsJ{{\em J.\ Phys.\ Soc.\ Japan }}
\def\ptps{{\em Prog.\ Theoret.\ Phys.\ Suppl.\ }}
\def\PTP{{\em Prog.\ Theoret.\ Phys.\  }}

\def\JMP{{\em J. Math.\ Phys.} }
\def\NPB{{\em Nucl.\ Phys.} B}
\def\NP{{\em Nucl.\ Phys.} }
\def\PLB{{\em Phys.\ Lett.} B}
\def\PL{{\em Phys.\ Lett.} }
\def\PRL{\em Phys.\ Rev.\ Lett. }
\def\PRB{{\em Phys.\ Rev.} B}
\def\PRD{{\em Phys.\ Rev.} D}
\def\PRe{{\em Phys.\ Rep.} }
\def\AP{{\em Ann.\ Phys.\ (N.Y.)} }
\def\RMP{{\
em Rev.\ Mod.\ Phys.} }
\def\ZPC{{\em Z.\ Phys.} C}
\def\SCI{\em Science}
\def\CMP{\em Comm.\ Math.\ Phys. }
\def\MPLA{{\em Mod.\ Phys.\ Lett.} A}
\def\IJMPB{{\em Int.\ J.\ Mod.\ Phys.} B}
\def\PR{{\em Phys.\ Rev.} }
\def\cmp{{\em Com.\ Math.\ Phys.}}
\def\JPA{{\em J.\  Phys.} A}
\def\CQG{\em Class.\ Quant.\ Grav. }
\def\ATMP{{\em Adv.\ Theoret.\ Math.\ Phys.} }
\def\ibid{{\em ibid.} }

\leftline{\bf References}

\renewenvironment{thebibliography}[1]
        {\begin{list}{[$\,$\arabic{enumi}$\,$]}  
        {\usecounter{enumi}\setlength{\parsep}{0pt}
         \setlength{\itemsep}{0pt}  \renewcommand{\baselinestretch}{1.2}
         \settowidth
        {\labelwidth}{#1 ~ ~}\sloppy}}{\end{list}}

\end{document}